\documentclass[twocolumn,showpacs,preprintnumbers,amsmath,amssymb]{revtex4}

\begin{document}
\draft

\title{non-ergodic mesoscopic systems}

\author{P. Singha Deo}
\address{Unit for Nano Science and Technology,
S. N. Bose National Centre for Basic Sciences, JD Block,
Sector III, Salt Lake City, Kolkata 98, India.}
\date{\today}

\begin{abstract}
Suppose there is a mesoscopic system connected to single channel leads.
If the system is non-chaotic or non-ergodic then the thermodynamic and
transport properties do not depend on 
impurity averaged density of states. 
We show that the partial 
density of states as well as density of states of a given system
can be determined exactly from the asymptotic wave-function 
(or scattering matrix)
at the resonances. The asymptotic wave-function can be determined
experimentally without any knowledge about the quantum mechanical
potential (including electron-electron interaction) or wave function
in the interior of the system. 
Some counter intuitive relations derived here can allow this.
\end{abstract}


\maketitle

For bulk samples,
ensamble averaging makes it un-necessary to know the exact
impurity configuration
and the exact Hamiltonian
of the sample. This is because experiments can only
observe ensamble averaged physical properties.
Ensamble averaging works because
of the well known ergodicity
hypothesis and the fact that sample to sample fluctuation is
not very large \cite{hua}.
For mesoscopic systems even when dealing with an ensamble,
sample to sample fluctuations are often so large that we cannot
talk of any averaged physical quantity \cite{dat}.
If disorder averaging cannot be taken then it becomes
necessary to know the exact impurity configuration of a
system, a seemingly impossible task for an experimentalist.
Also sometimes for mesoscopic systems
ergodicity itself does not hold for a single sample
that is further illustrated below.
In this work we show some exact relations.
This relation is not only counter intuitive but can also
provide a way to experimentalists, of bypassing the knowledge of
the exact Hamiltonian of the system to arrive at thermodynamic
and transport properties of mesoscopic system.

Consider an arbitrary potential $V(x,y)$ in the shaded region $\Omega$ of
Fig. 1.
The $z$ degree of freedom is usually frozen due to strong confinement
in the $z$ direction.
This potential can define a mesoscopic system that could
be a quantum dot or a quantum ring or anything else \cite{bay}.
Typically such a system is coupled to leads or measuring probes.
The leads connect the system
to reservoirs that are at fixed chemical potentials.
The reservoirs inject (or absorb) electrons to (or from) the
system through the leads.
We consider single channel leads as most experiments are done
with single channel leads \cite{dat}.
When leads are multichannel, then we do not have enough control
over the properties of the system to make it of practical use \cite{dat}.
By taking specific forms of $V(x,y)$ we can make the system to leads
coupling to be weak or strong.
For a fully chaotic
(or ergodic)
system, electrons will access all the states in the system
specially when 
the coupling to the leads is very weak and the electron spends
sufficient time in the system. However, for a non-chaotic (or non-ergodic)
system, all the states will not be accessed. Only part of the states will
be accessed that will depend on the position and the details of the leads
and the system, that will constitute the partial density of states (PDOS).
So the contribution of these electrons to thermodynamic observables
like quantum capacitance of the system, or heat capacity of the system,
as well as the linear response non-equilibrium effects will be
determined by this PDOS.
This PDOS cannot be determined from the Hamiltonian of the
isolated system as the PDOS depends on initial conditions
(that is through which lead the electrons enter the system and
through which lead they leave as well as the characteristics
of the leads and the system).
However, the scattering
matrix depends on these factors.
So the scattering matrix has more information than the Hamiltonian
and scattering matrix formulation is very important for mesoscopic systems
\cite{but86}.

The many body Hamiltonian for a N particle system is a
function of 3N coordinates. It has been proved that
the motion of one of these N
particles is governed by an effective potential that is just
a one body potential and a function of 3 coordinates \cite{koh}.
So in the preceding paragraph, when we refer to the
confining potential $V(x,y)$ of the system, we are referring to this
one body effective potential that includes electron-electron
interaction exactly.
The determination
of this one body potential is however very difficult and has
never been exact up to date \cite{man}.
For bulk systems there are fairly
good although approximate methods for obtaining the effective potential,
but the same need not be applicable to finite systems.
So for mesoscopic systems if we can bypass the determination
of the internal details and the exact Hamiltonian of the system,
by using the S-matrix then the S-matrix includes the effect
of electron-electron interaction exactly.

The approach proposed in this paper is due to some recent experiments
\cite{sch,yah} that are motivated by the possibility of obtaining
important
information from the scattering phase shift. So the present
work is an effort to identify information that can be obtained
from such experiments.
A series of experiments \cite{kob1, kob2} also tell us that
resonances in such a system as that schematically shown in Fig. 1
are Fano resonances. Recently it has been shown for some particular
potential (namely a delta function potential) in a quantum wire,
that at the Fano resonance
the density of states (DOS) and partial density of states
(PDOS) can be determined exactly from semi-classical formulas
involving scattering phase shift,
although Fano resonance is a purely quantum phenomenon
\cite{swa}. In this work we show that these results are true
for any general potential $V(x,y)$.

We shall use the following identity to arrive at our conclusions.
It can be rigorously derived for one dimension (1D) that is
naturally also true for quasi one dimension (Q1D)
\cite{aro} that involves tracing over sub-band index. In simpler form,
\cite{gra} for any such system schematically shown in Fig. 1.
\begin{equation}
-\int_{\Omega} d^3r \Lambda^*_{mn} {\delta \Lambda_{mn} 
\over \delta V(r)} = \Lambda^*_{mn} 
{d\Lambda_{mn} \over dE_m} + {1 \over 4E_m} 
(\Lambda_{mn}-
\Lambda^*_{mn})
\end{equation}
Here $\Lambda$ denotes scattering matrix and $E_m$ denotes the
kinetic energy in the $m$th channel that is further defined below.

The Schrodinger Eq. describing the system is
\begin{equation}
-{\hbar^2 \over 2 \mu}({\partial^2 \over \partial x^2} +
{\partial^2 \over \partial y^2}) \psi(x,y) + V(x,y) \psi(x,y) = E \psi(x,y)
\end{equation}
For $|x| < a$, $V(x,y)$ is negative enough to create at least one 
bound state (otherwise there will be no Fano resonance).
For $|x|\ge a$, $V(x,y)=V_c(y)$, where $V_c(y)$ is the confinement potential
in the leads.
The most general 
solution to Eq. 2 in different regions of 
Fig. 1 is
\begin{equation}
\psi^{(L)}(x,y)={e^{ik_mx} \over \sqrt{k_m}} \theta_m(y) 
+\Sigma_{n=1}^{\infty} {e^{-ik_nx}\over \sqrt{k_n}} r_{nm}
\theta_n(y)  (x\le -a)
\end{equation}
\begin{equation}
\psi^{(R)}(x,y)=\Sigma_{n=1}^{\infty} {e^{ik_nx}\over \sqrt{k_n}} t_{nm}
\theta_n(y) (x\ge a)
\end{equation}
Here $\theta_n(y)$ is the solution to the following Eq. that holds for
$|x|\ge a$.
\begin{equation}
[-{\hbar^2 \over 2 \mu}{\partial^2 \over \partial y^2} + V_c(y)] 
\theta_n(y)= \epsilon_n \theta_n(y)
\end{equation}
Therefore,
\begin{equation}
E=\epsilon_n + {\hbar^2 k_n^2 \over 2 \mu}
=\epsilon_n + E_n
\end{equation}
If $\epsilon_n > E$ then one can see from Eq. 6 that $k_n$ is imaginary
(the mode is evanescent).
The electron is incident along the $m$th channel which implies
$\epsilon_m < E$. For all $n \le m$, $ \epsilon_n <E $ and such channels
are propagating.
For $n \ge (m+1)$,$ \epsilon_n >E $, implying that these
channels are evanescent.
For $n \le m$,
$r_{nm}$ and $t_{nm}$ gives reflection and transmission amplitudes,
respectively. They also constitute the elements of the scattering
matrix $\Lambda$. For $n>m$, $t_{nm}$ and $r_{nm}$ gives transition
amplitude from the propagating mode to the $n$th evanescent mode
in the right lead and left lead, respectively.
However, they are not scattering matrix elements.

We can now define two functions $\psi^{(ev)}(x,y)$ and 
$\psi^{(od)}(x,y)$
such that $\psi^{(ev)}(x,y)=\psi^{(ev)}(-x,y)$ and 
$\psi^{(od)}(x,y)=-\psi^{(od)}(-x,y)$.
\begin{equation}
\psi^{(ev)}(x,y)=\Sigma_{n=1}^{\infty}(\delta_{nm}e^{-ik_nx}
-S^{(ev)}_{nm}e^{ik_nx}) 
{\theta_n(y)\over \sqrt{k_n}}
\end{equation}
\begin{equation}
\psi^{(od)}(x,y)=\Sigma_{n=1}^{\infty}(\delta_{nm}e^{-ik_nx}
-S^{(od)}_{nm}e^{ik_nx}) 
{\theta_n(y)\over \sqrt{k_n}}
\end{equation}
Then one can see that $\psi^{(L)}$ as well as $\psi^{(R)}$ is
given by
\begin{equation}
{1 \over 2} (\psi^{(ev)}-\psi^{(od)})
\end{equation}
where 
\begin{equation}
r_{nm} = - (S^{(od)}_{nm} + S^{(ev)}_{nm})/2
\end{equation}
\begin{equation}
t_{nm} =  (S^{(od)}_{nm} - S^{(ev)}_{nm})/2
\end{equation}
This works because any function can be written as a sum of an even
function and an odd function. And any square matrix can be written
as a sum of a symmetric matrix and an antisymmetric matrix.

Due to the same principle,
the wave-function in the scattering region $\Omega$ can be written as a sum
of an even function and an odd function. We denote them as
$\phi^{(ev)}_n(x,y)$ and $\phi^{(od)}_n(x,y)$.
\begin{equation}
\phi^{(ev)}_n(x,y)=\Sigma_{m=1}^\infty c_m \chi^{(ev)}_m(x,y)
\end{equation}
\begin{equation}
\phi^{(od)}_n(x,y)=\Sigma_{m=1}^\infty c_m \chi^{(od)}_m(x,y)
\end{equation}
where $\chi^{(ev)}$ and $\chi^{(od)}$ satisfy the Schrodinger
Eq. 2 with $V(x,y)=0$ in the region $\Omega$.
Therefore $\chi^{(ev)}$ and $\chi^{(od)}$ will satisfy the following
Eqs.
\begin{equation}
\chi^{(eo)}_m(a,y)=\theta_m(y) \,\, for \,\, |y| \le b
\end{equation}
\begin{equation}
\chi^{(eo)}_m(x,y)=0 \, \, for\, \,  y=F(x) \,\, or \,\, G(x)
\end{equation}
\begin{equation}
\chi^{(ev)}_m(x,y)= \chi^{(ev)}_m(-x,y)
\end{equation}
\begin{equation}
\chi^{(od)}_m(x,y)= -\chi^{(od)}_m(-x,y)
\end{equation}
$F(x)$ and $G(x)$ define the two curves at the upper and lower
boundaries of the region $\Omega$.
Here `$eo$' stands for `$ev$ or $od$'.
We also define the following matrix elements
\begin{equation}
F^{(eo)}_{n,m}={1 \over b(k_m k_n)^{1\over 2}}
\int_{-b}^{b} \theta_n(y) ({\partial \chi^{(eo)}_m 
\over \partial x})_{x=a} dy
\end{equation}

Now we require $\phi^{(eo)}_n$ and ${\partial \phi^{(eo)}_n
\over \partial x}$ to be continuous at $x=a$ for all $|y| \le b$.
Thus we get
\begin{equation}
\Sigma_{m=1}^\infty(\delta_{n,m}e^{-ik_m a} - S^{(eo)}_{m,n}e^{i k_m a})
{\theta_m(y) \over \sqrt{k_m}} =
\Sigma_{m=1}^{\infty} c_m \theta_m(y)
\end{equation}
$$-\Sigma_{n=1}^\infty i \sqrt{k_m}(\delta_{n,m}e^{-ik_m a} + 
S^{(eo)}_{m,n}e^{i k_m a})
\theta_m(y)  =$$
\begin{equation}
\Sigma_{m=1}^{\infty} c_m ({\partial \chi^{(eo)}_m \over \partial x})_{x
=a}
\end{equation}
Multiplying Eq. 19 and Eq. 20 with ${1 \over b} \theta_n(y)$
and integrating from $y=-b$ to $y=b$ and then combining them
we get a single matrix equation.
\begin{equation}
\Sigma_{m=1}^\infty (F^{(eo)}_{qm} - i \delta_{qm})e^{i k_m a}
S^{(eo)}_{mn}=(F^{(eo)}_{qn} + i \delta_{qn})e^{-ik_n a}
\end{equation}
or
\begin{equation}
S^{(eo)}_{mn}=e^{-ik_ma}[1+2i(F^{(eo)}-i1)^{-1}]_{mn}e^{-ik_n a}
\end{equation}

It is known in scattering theory that the bound states of the potential
$V(x,y)$ can also be obtained from Eqs. 7 and 8 by omitting the terms
$\delta_{nm}e^{ik_n x}$. Without this term Eqs. 7 and 8 are solutions
to the Sc. Eq. in 2 with correct boundary condition wherein there is
no incident wave. Identical analysis that lead to Eq. 21 in this case
gives
\begin{equation}
\Sigma_{m=mt}^{\infty} [F^{(eo)}_{qm} - i \delta_{qm}] e^{-\kappa_m a}
S_{mn}^{(eo)} =0
\end{equation}
Here $mt$ is the threshold value of $m$ for which bound states exist.
And $\kappa_m = i k_m$.
For $m<mt$ states will be scattering states. 
Supposing only the first channel is propagating then $mt=2$.
This will be further
illustrated soon. Solutions to Eq. 23 or
solutions to the following Eq. will give bound states.
\begin{equation}
det[F^{eo}_{cc} -i 1] = 0
\end{equation}
Here `$cc$' means closed channel.
Let us partition $F^{(eo)}$ into propagating and evanescent 
(or closed) channels.
\begin{eqnarray}
F^{(eo)} &=& \left (\begin{array}{cc}
F^{(eo)}_{pp} & F^{(eo)}_{pc}\\
F^{(eo)}_{cp} & F^{(eo)}_{cc}
\end{array} \right)
\end{eqnarray}
Therefore,
\begin{eqnarray}
\left ( \begin{array}{cc}
F^{(eo)}_{pp}-i1 & F^{(eo)}_{pc} \\ F^{(oe)}_{cp} &
F^{(eo)}_{cc}-i1
\end{array} \right ) 
\left ( \begin{array}{cc}
F^{(eo)}_{pp}-i1 & F^{(eo)}_{pc} \\ F^{(oe)}_{cp} &
F^{(eo)}_{cc}-i1
\end{array} \right )^{-1} 
\end{eqnarray}
$$= 
\left ( \begin{array}{cc}
1 & 0 \\ 0 & 1
\end{array} \right )$$ 
From this one can show that
\begin{equation}
[(F^{(eo)} -i1)^{-1}]_{pp}=[F^{(eo)}_{pp} - i1 - F^{(eo)}_{pc}(F^{(eo)}_{cc} 
- i1)^{-1}F^{(eo)}_{cp}]^{-1}
\end{equation}

So from Eq. 22 and Eq. 27 (for $m$ and $n$ being propagating channels)
$$S^{(eo)}_{mn}=$$
$$e^{-ik_ma}[1 + 2i[F^{(eo)}_{pp} - F^{(eo)}_{pc}
(F^{(eo)}_{cc}-i1)^{-1}
F^{(eo)}_{cp} - i1]^{-1}
]_{mn}e^{-ik_na}$$
\begin{equation}
= e^{-ik_ma}[(G^{(eo)}-i1)^{-1}(G^{(eo)}+i1)]_{mn} e^{-ik_na}
\end{equation}
where
\begin{equation}
G^{(eo)}_{mn}=[F^{(eo)}_{pp}-F^{(eo)}_{pc}(F^{(eo)}_{cc}-i1)^{-1}
F^{(eo)}_{cp}]_{mn}
\end{equation}

If there is only one propagating 
channel then $m=n=1$.
Also for one propagating channel, $p=1$.
Hence $G^{(eo)}$ becomes a number.
Therefore from Eq. 28
\begin{equation}
S^{(eo)}_{11}=e^{-2ik_1a}{G^{(eo)}+i \over G^{(eo)}-i}
=e^{2i(arccot(G^{(eo)}-k_1a)}
=e^{2i \delta^{(eo)}}
\end{equation}
where 
\begin{equation}
G^{(eo)}=F^{(eo)}_{11}-\Sigma_{m=2, n=2} F^{(eo)}_{1m} 
[(F^{(eo)}_{cc} -i1)^{-1} ]_{mn} F^{(eo)}_{n1}
\end{equation}
and 
\begin{equation}
\delta^{(eo)}=arccot(G^{(eo)}-k_1a)
\end{equation}
All the infinite $S_{mn}$ appear in $S_{11}$ through
$F_{cc}$, which carry information
of the entire Hilbert space accessed by the incident wave.
For non-ergodic systems there can be many more states
not accessed by the incident wave, ie., the transition amplitude
between these states and the incident wave being 0. These
states are not populated. So they neither contribute to
DOS and PDOS nor to the scattering matrix.
Trivial examples are cases like when the incident wave has a certain
symmetry that is incompatible with the symmetry of
these states. More generally,
there can be localized and scarred states that are known to occupy
a subset of the Hilbert space and do not connect to the rest.

Threshold energy $E$ for 1st closed channel is 
given by ${2 \mu \over \hbar^2}(E-\epsilon_2)>0$.
Below this energy the 2nd channel can have bound states. Such bound states
will occur at energies given by the solution to Eq. 24.
At these energies the 1st channel will be
propagating as its threshold is 
given by ${2 \mu \over \hbar^2}(E-\epsilon_1)>0$ 
and $S_{11}$ is given by Eq. 30. But at these energies $G^{eo}$
will diverge as it includes matrix elements of $[F^{eo}_{cc}-i1]^{-1}$
as can be seen from Eqs. 31 and 24. That in turn implies that
at a Fano resonance (as can be seen from Eq. 32)
\begin{equation}
\delta^{(ev)}=m \pi \,\, and \,\,
\delta^{(od)}=n \pi
\end{equation}
This is consistent with the fact that at Fano resonance $t_{11}$=0
(which is the definition of Fano resonance) as can be seen from
Eqs. 11 and 33.
Therefore $t_{11}-t_{11}^*$=0 at Fano resonance.
Also from Eqs 10 and 33 at Fano resonance
\begin{equation}
r_{11}-r_{11}^*=-i[sin(2 \delta^{(e)}) + sin(2 \delta^{(o)}]=0
\end{equation}
Therefore, at Fano resonance
\begin{equation}
\Lambda-\Lambda^\dagger=
\left ( \begin{array}{cc}
r_{11}-r_{11}^* & t_{11}-t_{11}^* \\ t_{11}-t_{11}^* & r_{11}^*-r_{11}
\end{array} \right )
=0
\end{equation}

Therefore at Fano resonance, from Eqs. 1 and 35
$$-{1 \over 4\pi i} \int_{\Omega} d^3r \Lambda^*_{mn} 
{\delta \Lambda_{mn} \over \delta V(r)} - HC 
={1 \over 4\pi i}  (\Lambda^*_{mn} 
{d\Lambda_{mn} \over dE} - HC) $$
\begin{equation}
={1 \over 2 \pi} |\Lambda_{mn}|^2
{d[Arg(\Lambda_{mn})] \over dE}
\end{equation}
Here $Arg(\Lambda_{mn})=Arctan{Im\Lambda_{mn}
\over Re\Lambda_{mn}}$
The complicated integral involving the local potential on LHS is
the PDOS for an electron incident in channel $n$ and scattered
to channel $m$. So at the Fano resonance
the PDOS can be determined exactly from the scattering
phase shift ${d[Arg(\Lambda_{mn})] \over dE}$ 
and $|\Lambda_{mn}|^2$.
Both these quantities were measured in \cite{sch, kob1, kob2}.
Summing over $m$ and $n$ we get at Fano resonance
\begin{equation}
-{1\over 4\pi i}\Sigma_{mn}
\int_{\Omega} d^3r \Lambda^*_{mn} 
{\delta \Lambda_{mn} \over \delta V(r)} - HC 
= {d[{1\over 2\pi i} log Det[\Lambda]] \over dE}
\end{equation}
which is Friedel sum rule. It is expected to hold good only for bulk
samples that are in semiclassical limit. For finite systems that are
in quantum regime there is always a correction term arising from
${1 \over 4E_m}(\Lambda_{mn} -\Lambda^*_{mn})$. 
We do not know of any example where
this correction term becomes exactly 0. In Q1D, although Fano resonance is
a purely quantum interference phenomenon,
the correction terms are exactly 0 making Friedel
sum rule exact. The correction terms are extremely non-universal
(resonances are generally characterized by line shape i.e., $|\Lambda_{
mn}|^2$ and scattering phase shifts $Arg(\Lambda_{mn})$)
and also depend on sample specific parameters (e.g., $E_m$ depends on
$V_c(y)$ as well as material parameters like bottom of conduction band,
effective mass $\mu$, etc.)

In the experiments of Refs. \cite{sch,kob1,kob2}
${d\over dE}Arg(t)$ turns out to be negative at the Fano
resonance.
This means that PDOS at a Fano resonance is negative.
That in turn means that the quantum capacitance at a Fano resonance
can be negative. Which means the effective potential ($V=Q/C$) due to
some negatively charged electron in a quantum dot at a Fano
resonance can be positive. So negatively charged electrons
in the leads
will be attracted or in other words there will be electron-electron
attraction at a Fano resonance of a quantum dot. Such an
attraction was observed in numerical simulations
\cite{deo}, although
a proper explanation could not be given.

As a simple experiment one can 
repeat the experiments of Ref. \cite{sch, kob1, kob2} with
distorted quantum dots for which the spacing between the
peaks in $|\Lambda_{mn}|^2$ are not expected to be uniform.
This spacing will depend on DOS which for distorted dots
do not peak at regular intervals. Thus from the spacings one can
get the LHS of Eq. 36. Also from similar data as that in Refs.
\cite{sch,kob1,kob2} one can get RHS of Eq. 36 as explained before.

In conclusion, Friedel sum rule and similar semiclassical formulas
(Eqs. 36 and 37)
become exact at resonances for any general potential in single
channel Q1D that can support a resonance. This is very
counter intuitive as Fano resonance is a purely quantum phenomenon.
We do not know of any other situation where semiclassical formula
can become exact as mesoscopic world is always quantum. Semiclassical
formula can at most be a good approximation.
Thereby, the experimental data of Refs. \cite{sch,kob1,kob2} do
carry important information. Using such data 
experimentalists can bypass the knowledge of
microscopic Hamiltonian to know the DOS and PDOS and hence 
thermodynamic and transport properties.

\end{document}